\documentclass{basi}
\usepackage{graphicx}
\begin{document}
\title[Photometry of HD 40372]{Photometry of the $\delta$ Scuti star HD 40372 }
\author[Deb et al.]%
       {Sukanta Deb$^1$ \thanks{e-mail: sdeb@physics.du.ac.in},
       S. K. Tiwari$^2$, Harinder P.
Singh$^1$, T. R. Seshadri$^1$, U. S. Chaubey$^2$
       \\
 $^1$ Department of Physics \& Astrophysics, University of Delhi, Delhi 110007, India\\
 $^2$  Aryabhatta Research Institute of Observational Sciences (ARIES), Manora Peak, Nainital 263129, India}

%\date{Received 2001 May 30; accepted 2001 June 07}

\maketitle
\label{firstpage}
\begin{abstract}
We present B band photometry of the $\delta$ Scuti star HD 40372
using the ARIES three channel fast photometer attached to the
$104$-cm Sampurnanand telescope in high-speed photometric mode. The
star was observed for $\sim$ 5 hours on December 13, 2008. Based on
the high quality photometric data, we have done period analysis by
various periodogram analysis techniques. The best estimate of the period 
is found to be $\sim 0.067$ days. With this period and the other stellar 
parameters determined from $uvby\beta$ photometry available in the literature, 
we have calculated the $Q$ value for the star. Comparison of this $Q$ value
with that determined from the model calculations shows that the star
is pulsating in p$_{3}$ mode with $l = 2$.
\end{abstract}
\section{Introduction}
\label{sec:intro} Pulsating stars play an important role in probing
the physical properties of stars, including their masses,
luminosities, temperatures and metallicities. Pulsating stars exist
over a broad range of stellar parameters and evolutionary stages.
The existence of pulsating stars as components of binary systems has
drawn a considerable attention in the recent past because binary
stars provide useful information on the physical properties of the
components such as masses and radii. In addition, the pulsating
component offers an independent way of estimating its parameters
such as the pulsation constant which is useful for identification of
modes (Derekas et al. 2009).

$\delta$ Scuti stars are short period pulsating variables located in
the lower part of the classical instability strip on or above zero
age main sequence (ZAMS) in the Hertzsprung-Russell (HR) diagram with
masses ranging from 1.0 to 3.0 M$_\odot$. The typical time scales of
these oscillations range from few minutes to some hours (from 18 min
to 8 h) (Breger 2000). Since the time scale of oscillations of these
class of stars is short and they are bright, the time sampling of
the light curves is of the order of few tens of seconds and requires
fast photometry.

These variables are spectral type A to F dwarfs or sub giants that
have not yet reached the red giant branch and hence are mainly
moderately evolved objects (Alcock et al. 2000). The majority of
these variables pulsate with very low amplitudes, ranging from 0.01
mag to several tenths of a mag in V  with non-radial, low-degree
($l$ $\leq$  3) and low order (k = 0 to 6) p modes (Dawson et al.
1995). Some are radial pulsators  and some others also pulsate in a
mixture of radial and non-radial modes (Breger et al. 1999). A list
of pulsating $\delta$ Scuti stars in stellar systems can be found in
Lampens \& Boffin (2000). More recent complete list of $\delta$
Scuti stars detected in eclipsing binary systems are given in
Rodr\'{i}guez et al. (2004).

HD 40372 has a V magnitude of 5.904 mag and is located at $\alpha$ =
05$^{\rm h}$\,58$^{\rm m}$\,24$^{\rm s}$.44 and $\delta$ = +01$^{\rm o}$\,50$^{\rm '}$\,13$^{\rm ''}$.59 (Epoch J2000). The star has a
spectral type A5me. The Str\H{o}mgen indices of HD 40372 are $b-y = 0.122, 
m_{1} = 0.211, c_{1} = 0.960$ and H$_\beta$ = 2.806 (Hauck \& Mermilliod 1998)
. The star belongs to a spectroscopic binary system with orbital period of
2.74050 days (Nadeau 1951). Also, there are light curve data from
Hipparcos photometry with the listed orbital period of 2.74068
days. One of the components of the star HD 40372 was discovered to be a 
$\delta$ Scuti star by Breger (1973) having a pulsation period of 0.054 days. 
In a study of this star, Liu (1999) reported a photometric pulsation period of
0.062 days (16.461 c/d) and amplitude of 10.93 mmag in the
Str\H{o}mgen $v$ filter. Their period determination is not so
accurate and does not fit the light curve data well. On the other
hand, a period of 0.061 day and and an amplitude of 0.02 (V band)
has been catalogued in Rodr\'{i}guez and Breger (2001).

In this paper, we provide highly accurate photometric B band
out-of-eclipse data of the pulsating component of HD 40372. The
star was observed for a duration of $\sim$ 5 hours in highly
photometric sky conditions. In Section 2, we describe the
observation and data reduction procedure. Section 3 deals with the
analysis of the light curve. In Section 4, we derive the various
physical parameters of the star from the available $uvby\beta$
photometry in the literature and the pulsational constant. In
Section 5, we present the conclusions of our study.

\section{Observation and data reduction}
The star HD 40372 was observed in a clear sky condition on December
13, 2008 using the three-channel fast photometer attached to the
$\rm 1.04$-m Sampurnanand telescope at ARIES in an out-of-eclipse
phase. Details of the photometer are available in Ashoka et al.
(2001) and Gupta et al. (2001). The time series photometric observations
consist of continuous 10s integrations through a Johnson B filter.
An aperture of 30 arc-second is used to minimize the flux variations
caused by seeing fluctuations and tracking drift. Large diaphragm is
also used in order to diminish the noise caused by the wings of the
stellar image. Manual guiding after every 2 to 3 minutes ensured
high quality of centering of the star in the diaphragm. The sky
background is measured in a non-regular fashion to avoid periodic
break in the light curve. The sky background is measured by
interpolating the points with a piece wise linear function. The raw
data are corrected for the dead time counts (T = 23ns), sky
background and mean atmospheric extinction. The bad data points of
the light curve are visually inspected and removed during the data
reduction process. Finally, the data sets are expressed as
Heliocentric Julian Day (HJD) versus fractional magnitude with
respect to the mean magnitude of the light curve.\footnote{The
observational data presented in this paper can be obtained from the
authors on request.} The processed light curve is shown in the upper panel 
of Fig. 1.

\section{Analysis of the light curve}
\subsection{Detrending the data}
From Fig.~1 (upper panel), it is clearly seen that there is a slow trend 
present in the data. This trend in the data arises mainly due to the change in 
the sky transparency and/or change in the air-mass during the time of
observations and may be because of the use of single channel of the
photometer. In general, single channel photometry is not ideal for
measuring the brightness changes of variable stars. However for
bright and short period variable stars, such as $\delta$ Scuti
stars, under good photometric sky conditions, this technique of
measuring the magnitude changes can be shown to be adequate. In a
work by Joshi et al. (2003), it was in fact seen that single channel
photometry can be used for measuring light variations of short
period variable stars in favourable sky conditions.

The processed light curve data are detrended by subtracting out the slow
trends using a second order polynomial fit of the form
$a+bt+ct^{2}$, where $t$ is the time of observations and $a,
b, c$ are the coefficients determined from the fit. The procedure
removes the slow trends in the data. In the lower panel of Fig.~1,
we show the detrended data.
%%%%%%%%%%%%%%
%Figure 1
\begin{figure}
\begin{center}
\includegraphics[height=8cm,width=10cm]{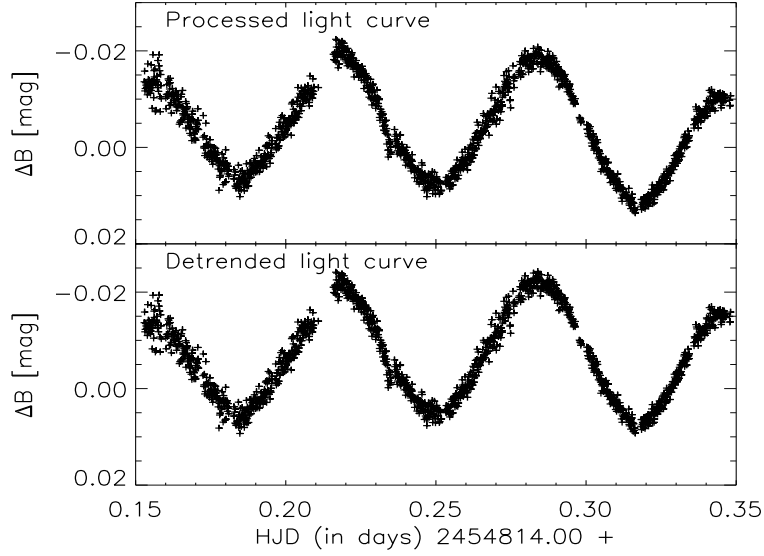}
\caption{The upper panel shows the processed light curve data of HD 40372. 
The lower panel shows the detrended data.}
\end{center}
\end{figure}

\subsection {Period determination}

There are a number of period determination techniques available for
determining the periodicity in a time series data. We have estimated
the period of HD 40372 using a number of algorithms. The various
periodogram analysis techniques were applied in order to see the
consistency of the result. We have calculated the period of HD
40372 using the following algorithms:
\begin{itemize}
\item[(1)]{Period04 algorithm (Lenz \& Breger 2005)}
\item[(2)]{Discrete Fourier Transform (DFT) algorithm (Deeming 1975)}
\item[(3)]{Non-linear least square fit (NLSQ) method (Press et al. 1992)}
\item[(4)]{Lomb \& Scargle (LS) periodogram (Lomb 1976, Scargle 1982, Horne \& Baliunas 1986)}
\item[(5)]{CLEAN algorithm (Roberts et al. 1987)}
\item[(6)]{Phase dispersion minimization (PDM) technique (Stellingwerf 1978) }
\item[(7)]{Multi-harmonic ANOVA (MULTIH) algorithm  (Schwarzenberg-Czerny 1996)}
\item[(8)]{Minimization of information entropy (ME) method (Cincotta, M$\acute{\rm e}$ndez \& N$\acute{\rm u}\tilde {\rm n}$ez 1995)}.
\end{itemize}
%%%%%%%%%%%%%%%%%%%%
%Figure 2
\begin{figure}
\begin{center}
\includegraphics[height=8cm,width=9cm]{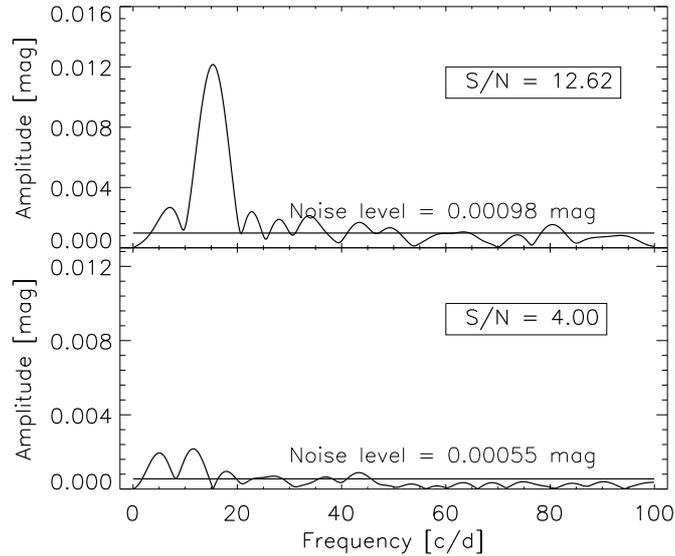}
\caption{The upper panel shows the amplitude spectrum of HD 40372 obtained
from Period04 for the frequency $f$ = 15.36492 c/d while the lower panel
shows the amplitude spectrum after subtracting the data corresponding to that
frequency.}
\end{center}
\end{figure}
For finding multiple frequencies in the time series data, the methods such as
Period04 and DFT are very much efficient. We have used Period04 method in
order to search for multiple periodicities. To search for multiple
periodicities in the data, the data corresponding to the sinusoid of the
first detected frequency are subtracted from the original data which is
generally called `pre-whitening' of the data. Period analysis is now carried
out on the residual data and the next frequency is identified. This procedure
is repeated until no significant frequencies are left in the data. In order to
search for statistically significant frequency in the time series data, we have
 used the criterion of Breger et al. (1993) that an amplitude signal to noise
ratio ($SNR$) should be greater than 4.0 to judge the reality of a
peak in the amplitude spectrum. It consists of calculation of the
$SNR$ of the frequency in the periodogram. This is done by finding
the highest amplitude of the peak for the frequency obtained and the
noise as the average amplitude in the residual periodogram in a
frequency range that encloses the detected peak after the detected
frequency is pre-whitened. In Fig.~2, we show the amplitude spectrum
of the HD 40372 obtained from Period04. The amplitude spectrum peaks
at 15.36492 $\pm$ 0.02106 c/d with amplitude 0.01232 $\pm$ 0.00010 mag 
(upper panel). Also
shown is the corresponding noise level of the peak. The noise level
is at 0.00098 mag yielding a $SNR$ of 12.620. The noise level is
calculated in the frequency range from [frequency-boxsize/2.0
,frequency+boxsize/2.0]. We have chosen boxsize = 20. We have also
detected another frequency at 11.52369 c/d (lower panel of
Fig.~2). SNR of this peak is $\sim$ 4.0. The reality of the peak is
suspicious as it lies on the boundary of the Breger's criterion and
because of the short span of observations. We check the presence of
the most dominant peak in the periodogram by means of other
periodogram analysis techniques: DFT, NLSQ, LS, CLEAN, PDM, MULTIH and ME 
methods. Since a frequency peak near  15.36492 c/d is found to be
present in all of the periodograms, it is taken to be the real and
most dominant frequency of the data.
\begin{table}
\caption{The dominant frequencies obtained from the eight different periodogram
analysis techniques. The final best estimate of the period is obtained by the
ME method.}
\begin{center}
\medskip
\label{Table1}
\begin{tabular}{lcccc}
\hline
Method   & $P$[days] & Amplitude [mag]$^{*}$ &$\sigma ^{*}$& $\Delta P$ \\ \hline
Period04 & 0.06508 & 0.01267 & 0.00248 & 0.00120 \\\hline
DFT      & 0.06539 & 0.01279 & 0.00237 & 0.00108 \\\hline
NLSQ     & 0.06782 & 0.01356 & 0.00208 & 0.00094 \\\hline
LS       & 0.06548 & 0.01283 & 0.00234 & 0.00100 \\\hline
CLEAN    & 0.06392 & 0.01221 & 0.00296 & 0.00140 \\\hline
PDM      & 0.06726 & 0.01344 & 0.00204 & 0.00094 \\\hline
MULTIH   & 0.06717 & 0.01342 & 0.00204 & 0.00089 \\\hline
ME       & 0.06732 & 0.01346 & 0.00204 & 0.00095 \\\hline
\end{tabular}
\end{center}
~~~~~~~*Determined from the Fourier fit to the phased light curve as 
described in Section 3.3. $\sigma^{2}$ is the variance of the noise 
after the signal has been subtracted, $\Delta P$  is the error in the period.
\end{table}
%%%%%%%%%%%%%%%%%
%Figure 3
\begin{figure}
\begin{center}
\includegraphics[height=8cm,width=9cm]{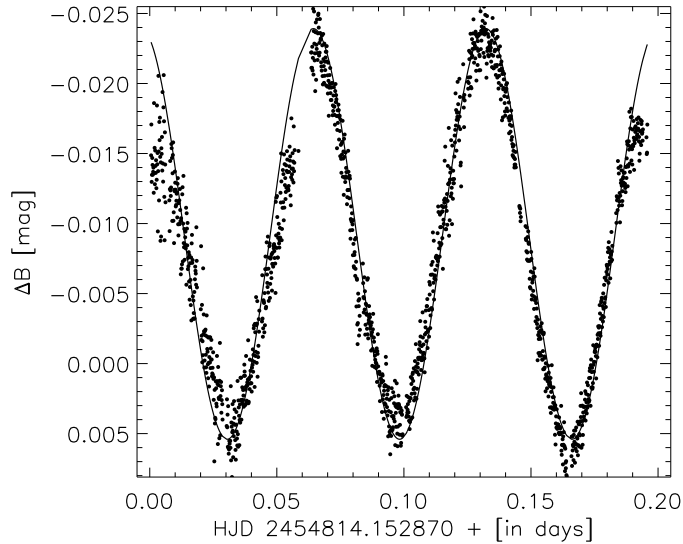}
\caption{A non-linear least square fit to the data yields a period of
0.06782 days (frequency 14.7454 c/d). The solid line is the best fit to the
data using NLSQ method.}
\end{center}
\end{figure}
%%%%%%%%%%%%%%%%
%figure 4
\begin{figure}
\begin{center}
\includegraphics[height=8cm,width=9cm]{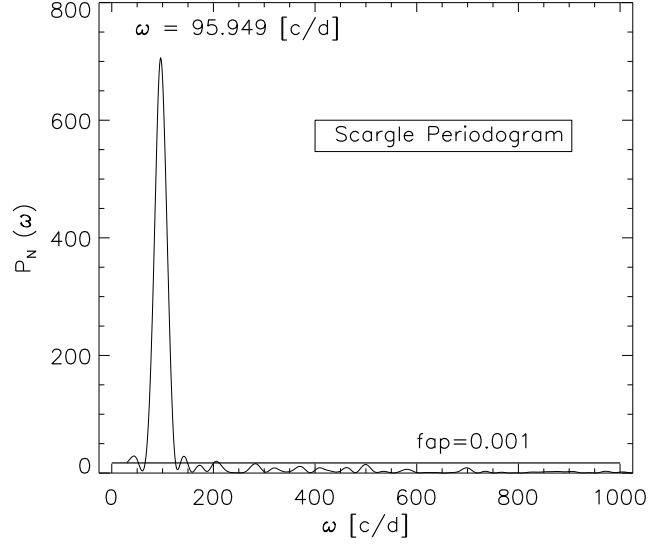}
\caption{Fourier transform of the data of HD 40372 using LS periodogram.
The periodogram was calculated in the range $f_{L}<f<f_{N}$, where the lowest
frequency (the longest period) that can be detected will be equal to the
length of the whole data string, namely $f_{L}=1/(t_{N_{obs}}-t_{1})$. The
highest frequency for equally spaced data is formally given by the Nyquist frequency, $f_{N}=1/(2\Delta t)$ , where $\Delta t$  is the sampling interval in the data string. $N_{obs}$ denotes the total number of data points, $t_{1}$ 
is the time of the start of observations.}
\end{center}
\end{figure}
%%%%%%%%%%%%%%%%%%%%%%
%Figure 5
\begin{figure}
\begin{center}
\includegraphics[height=10cm,width=12cm]{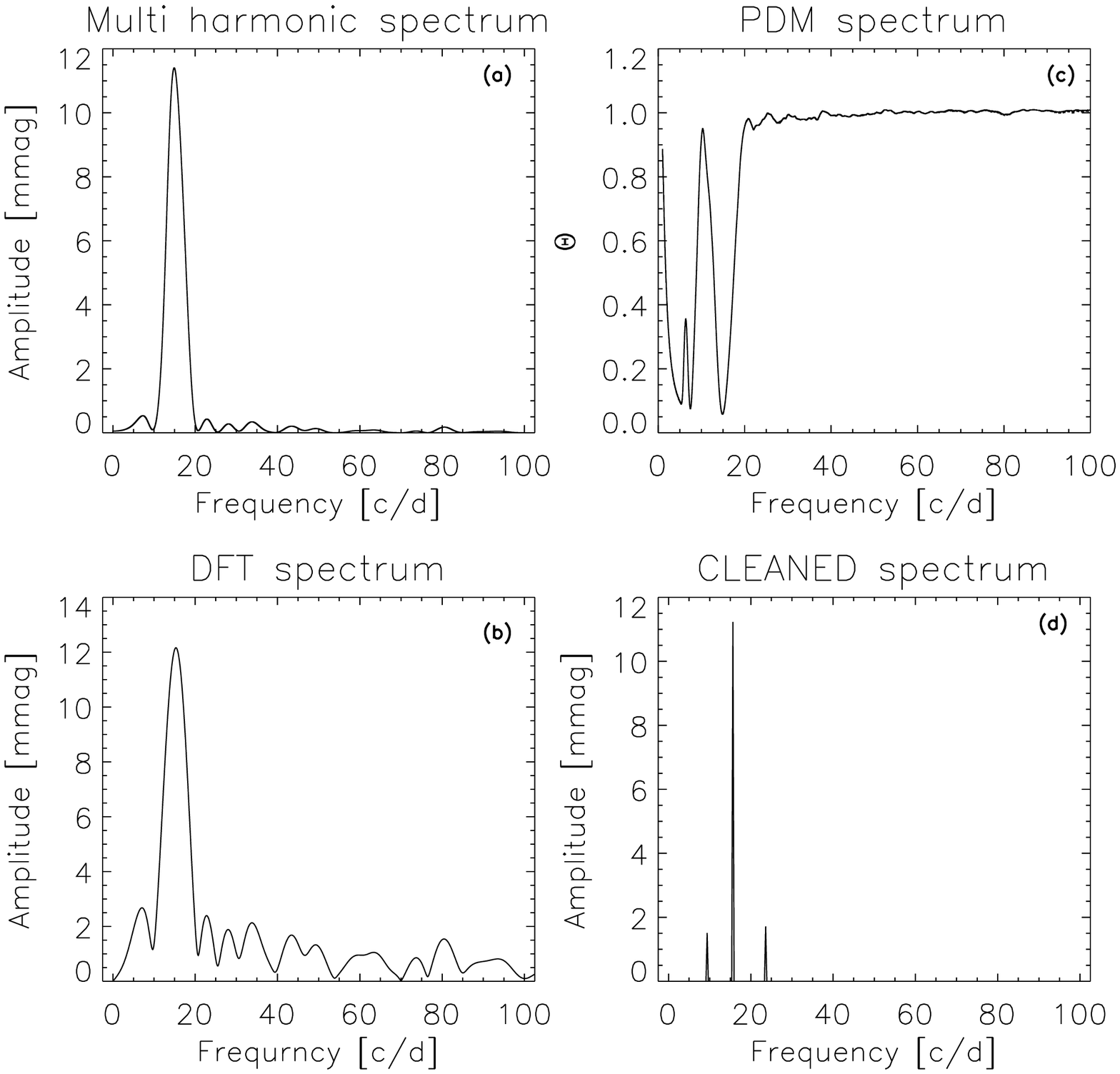}
\caption{Power spectra of HD 40372 obtained from MULTIH, DFT, PDM and CLEAN
are shown. The most dominant peak in all the spectra occurs near
to $\sim$ 15 c/d.}
\end{center}
\end{figure}

In Table 1 we list the period of HD 40372 as determined by the 
various methods mentioned above. The final best estimate of the
period $P = 0.06732$ days is obtained by the ME method. We search for the best 
period around 10\% of an approximate period as determined by the other
methods in steps of 10$^{-5}$ days to calculate the period correct
up to five decimal places. ME method, in principle, 
allows to compute periods with a higher numerical precision, since it is based 
on proper ordering of the light curve points in the phase-magnitude space with 
the true period corresponding to the minimum of the entropy. Also in comparison 
with other methods, this method is computationally straightforward and very 
fast, just requiring counting points within a binning scheme (Cincotta, M$\acute{\rm e}$ndez \& N$\acute{\rm u}\tilde {\rm n}$ez 1995). Methods such as ME and
string length (Dworetsky 1983) are, in general, used to improve the period 
determined by other techniques (cf. Derekas, Kiss \& Bedding 2007).

In Fig.~3, we plot the fit to the data with frequency 14.74540 c/d 
obtained from the NLSQ method. Fig.~4 shows the LS power spectrum with 
peak at $\omega = 95.94900 $ c/d corresponding to a frequency of  15.27076 c/d.
 In case of the Scargle periodogram (Fig.~4), two other peaks are detected 
with confidence level larger than 99\%. However, the existence of these 
multiple frequencies in the data needs to be confirmed by observations spanning 
over a longer time.
 
Fig.~5 shows the power spectra of HD 40372 obtained by MULTIH, DFT, PDM and 
CLEAN methods.  In Fig. 6, we plot the information entropy (S) versus Period 
($P$) for HD 40372 obtained from the ME method. In the left panels of Fig.~7, 
we plot the light curve of HD 40372 phased with the three periods $\sim$ 0.064,
 0.065 and 0.067 respectively. The residuals of the fitted light curves are 
shown in the corresponding right panels of Fig.~7. The $\sigma$ of the 
residuals is the least for $P \sim 0.067$ days which is taken to be the best 
period estimate.
%%%%%%%%%%%%%%%%%%%%%%%%
%Figure 6
\begin{figure}
\begin{center}
\includegraphics[height=10cm,width=12cm]{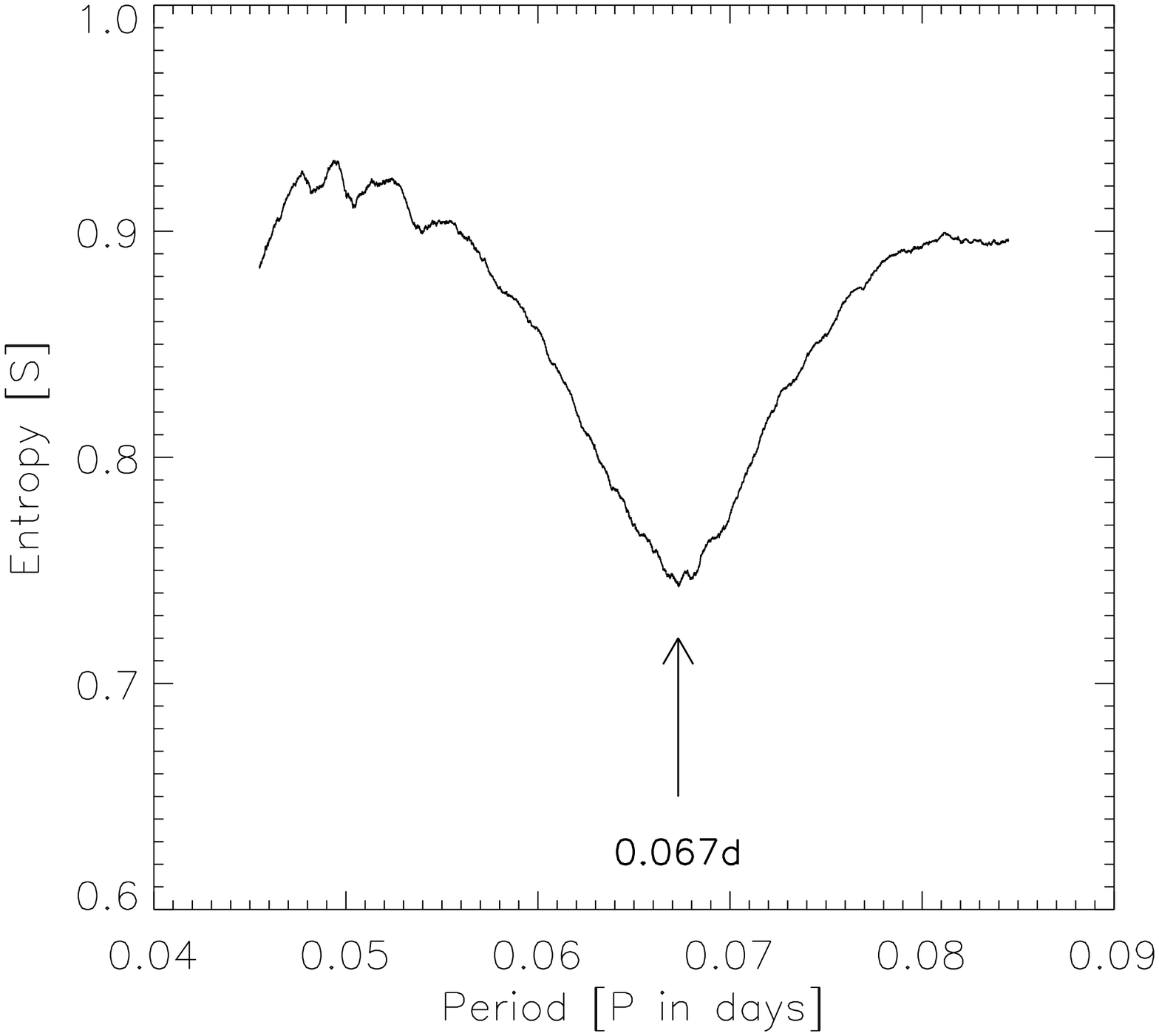}
\caption{Plot of information entropy (S) over a wide range of $P$ for 
the star HD 40372. S is minimum at a period $\sim$ 0.067 days.}
\end{center}
\end{figure}
%%%%%%%%%%%%%%%%%%%%
%Figure 7
\begin{figure}
\begin{center}
\includegraphics[height=11cm,width=12cm]{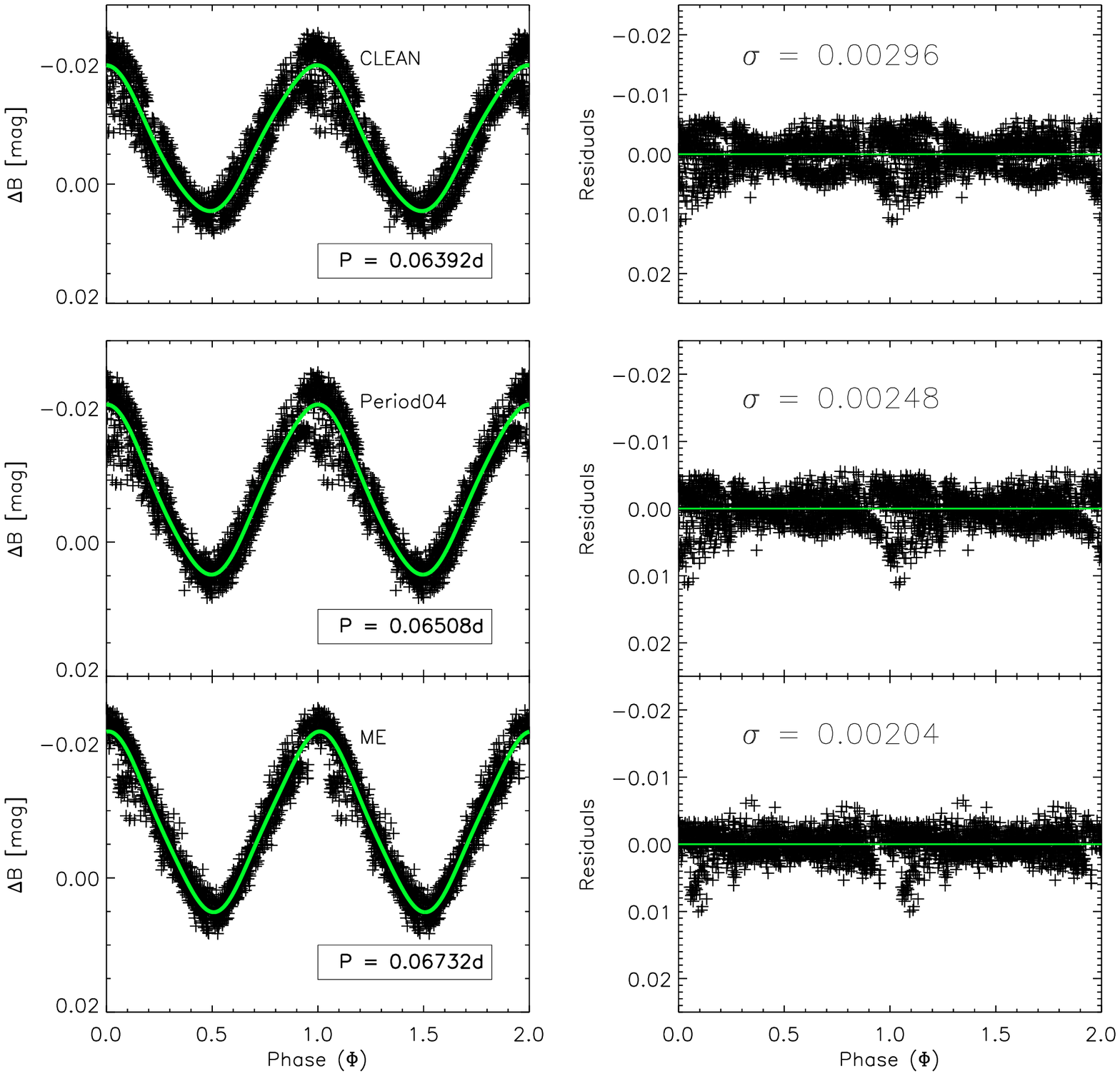}
\caption{The left panels show the light curve of HD 40372 phased with the 
three periods $\sim$ 0.064, 0.065 and 0.067 days respectively. The solid 
lines in the left panels are the fitted light curves. The light curve phased 
with these three periods consists of 1485 data points. The corresponding 
residuals from the fit and standard deviations of the residuals are shown in 
the right panels.}
\end{center}
\end{figure}
\subsection{Error Estimates}
Estimation of error ($\Delta P$) in the period determination is an important task to 
check the reliability of the period determined by various methods. As described
in the above, we have estimated period by the various techniques available in 
the literature. Some of the period determination techniques give nearly 
identical results, whereas, the others differ by an insignificant amount in 
the third decimal places. True and precise period is, in fact, the one which 
gives a smooth and well-ordered light curve when phased with it, whereas, other 
periods will give disordered and scattered light curves when phased with them.
Since the determination of errors by the various methods are different, we adopt
the following common technique for finding error in the period obtained from 
various techniques : 

Firstly, the light curve is phased with the respective periods determined by 
the various techniques. Then, individual phased light curves are fitted with a 
Fourier series of third order of the form as described in next
section so that the residuals consist of noise only. From each of the fitted 
light curves, amplitude of the signal (A), variance ($\sigma^2$) of the noise 
after the signal has been subtracted, are calculated. The error in the period is then calculated by the following formula (Kov\'{a}cs 1981, Horne \& Baliunas 1986):
\begin{eqnarray}
\Delta\omega = \frac{3\pi\sigma}{2N^{1/2}TA},
\end{eqnarray}
or
\begin{eqnarray}
\Delta P = \frac{3\sigma P^{2}}{4N^{1/2}TA},
\end{eqnarray}
  
 where $\omega$ is the angular frequency, $N$ is the number of data points, 
$T$ is the total length of the data. In table 1, we list the values of A, 
$\sigma$ of the residuals and the errors in the period obtained by the above 
technique. In the present case, $N = 1485$ and $T \sim 5 $ hours. From Table 1,
 It should be noted that the error in the period is least for period 
$\sim 0.067$ days.
\subsection{Pulsation parameters}

The pulsation parameter of HD 40372 is described by decomposing its
phased light curve in the following form :

\begin{equation}
m(\Phi) =  m_{0}+ \sum_{i=1}^{3}  a_{i}\cos(2\pi i \Phi(t))+ \sum_{i=1}^{3} b_{i}\sin(2\pi i \Phi(t)),
\end{equation}
where $m(\Phi$) is the phased light curve and $\Phi$ is defined by:
\begin{equation}
\Phi =\frac{\left( t-t_{0}\right) }{P}-Int\left( \frac{ t-t_{0}
}{P}\right).
\end{equation}
The value of $\Phi$ is from 0 to 1, corresponding to a full cycle of
pulsation and $Int$ denotes the integer part of the quantity. The
zero point of the phase corresponds to the time of maximum light
($t_{0}$). $m_{0}$ is the mean magnitude, $t$ is the time of
observations and $P$ is the period of the star in days. In Fig.~8,
we plot the phased light curve obtained with period 0.06732 days.
Points lying 2$\sigma$ away from the fit to the data are removed
using a multi-pass nonlinear fitting algorithm in IDL (Interactive
data language). The solid line is the 3rd order Fourier fit to the
outlier-removed data.

In Table 2, we list the Fourier parameters $a_{i}$'s and $b_{i}$'s
determined by using the period as estimated by the minimization of
entropy method.

\begin{table*}
\centering
\caption{Fourier parameters of the $\delta$ Scuti star HD 40372}
\scalebox{0.8}{
\begin{tabular}{lcccccccccc}
\\
\hline
\hline
ID & Period (in days) & $m_{0}$ & $a_{1}$ & $b_{1}$ & $a_{2}$ & $b_{2}$ & $a_{3}$ &$b_{3}$    \\
(1)&(2)&(3)&(4)&(5)&(6)&(7)&(8)&(9) \\
\hline
\hline
HD 40372 & 0.06732 & -0.00764&-0.01330&-0.00034&-0.00071&-0.00017&-0.00038&-0.00014 \\
& $\pm$0.00095&$\pm$0.00025&$\pm$0.00024&$\pm$0.00024&$\pm$0.00024&$\pm$0.00024&$\pm$0.00024&$\pm$0.00024\\ \hline
\hline
\end{tabular}
}
\end{table*}
%Figure 8
\begin{figure}
\begin{center}
\includegraphics[height=8cm,width=9cm]{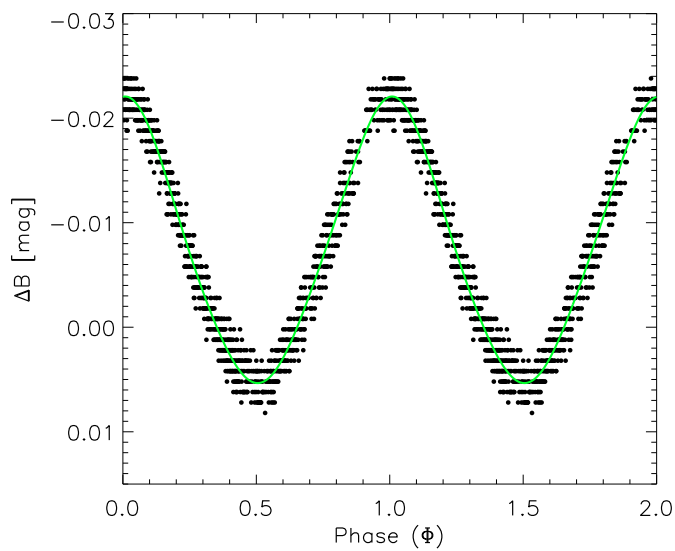}
\caption{Outlier-removed phased light curve of HD 40372 obtained from 
Eqn.~(4) with the period 0.06732 days as determined from the minimization of 
entropy method. The solid line is the 3rd order Fourier fit to the data.
}
\end{center}
\end{figure}

\section{Physical parameters and evolutionary status}

We make use of the $uvby\beta$ photometry from Hauck \& Mermilliod (1998)
to estimate some of the fundamental physical parameters of HD 40372
including absolute magnitude, the effective temperature, radius and
surface gravity. Using the model-atmosphere calibrations of $uvby\beta$
photometry by Moon $\&$ Dworetsky (1985) and later modified by Napiwotzki
 et al. (1993) , we obtain the following physical parameters T$_{eff}$ = 
7706\,$\pm \,30$\,K, M$_{v}$ = 1.23\,$\pm\, 0.19$, R = 2.78\,$\pm$\,0.21R
$_{\odot}$, $\log\,g = 3.71\,\pm0.06$. 
%%%%%%%%%%%%%%%%%%%%%
%Figure 9
\begin{figure}
\begin{center}
\includegraphics[height=8cm,width=9cm]{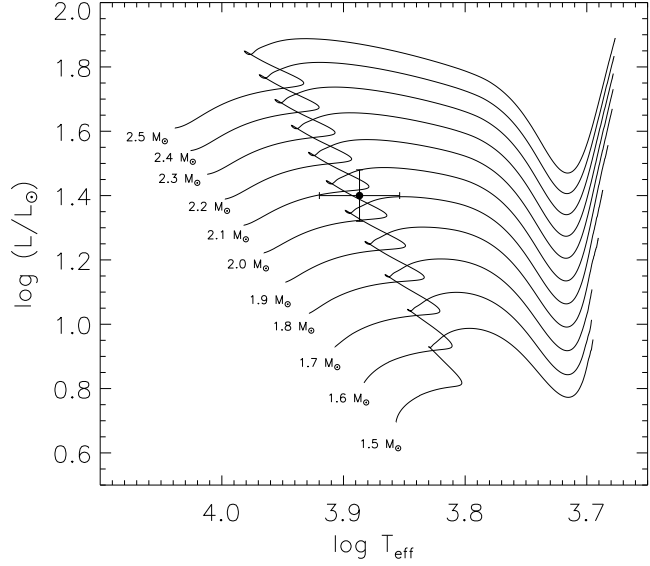}
\caption{HR diagram showing the position of HD 40372 (filled circle)
in the Cepheid instability strip. Evolutionary tracks for 1.5 to 2.5
M$_\odot$ from Christensen-Dalsgaard (1993) for composition X = 0.692827, 
Z = 0.02 are shown.}
\end{center}
\end{figure}

The mass can be calculated from Cox (1999):
\begin{eqnarray}
\log\,(M/M_{\odot}) = 0.46 - 0.10 M_{bol},
\end{eqnarray}

where M$_{bol} = M_{v} + BC$. A bolometric correction BC = 0.03 is calculated
from Flower (1996) corresponding to the derived temperature of 7706 K.
From the above equation we get an estimate of mass as M = 2.15 $\pm$ 0.02
M$_{\odot}$. 
In Fig.~9, we plot the position of HD 40372 in the HR diagram. The evolutionary
tracks\footnote{http://owww.phys.au.dk/$\sim$jcd/emdl94/eff\_v6/} of mass ranging from 1.5M$_{\odot}$ to 2.5M$_{\odot}$ are over-plotted
for composition X = 0.692827 and Z = 0.02 (Christensen-Dalsgaard 1993). 
We also estimate  mass from the evolutionary tracks. With T$_{eff}$
= 7706 $\pm$ 30 K and log (L/L$_ {\odot}$)= 1.40 $\pm$ 0.08 , we
derive an evolutionary mass of $\sim$ 2 M$_ {\odot}$ which is
comparable to the mass determined from Eqn.~(5).

The pulsational constant $Q$ of a star can be used to identify the pulsational
mode of a star. The $Q$ value of a pulsating star is given by the following
relation (Breger \& Bregman 1975):
\begin{eqnarray}
\log Q = -6.454 + \log P + 0.5 \log g +0.1 M_{bol} + \log T_{eff}.
\end{eqnarray}

Substituting the value of $P = 0.06732$ days with the calculated physical
parameters as described above leads to $Q$ = 0.01746 $\pm$ 0.00302.
In order to identify the pulsational mode of the star HD 40372, we compare
the observed $Q$ value with the theoretical calculations of Fitch (1981). The
2.5M65 model in the Fitch's Table 2C shows the possible solution with $l = 2$,
p$_{3}$ mode for the frequency pattern of HD 40372. However, the uncertainty
in the $Q$ value calculated from Eqn.~(4) is very large. According to Breger
(1990), $Q$ value calculation from the above relation contains an error of
18\%. Therefore it is very difficult to find mode accurately with $Q$ value alone. Simultaneous color measurements would be helpful to determine the 
mode of HD 40372 more accurately.

\section{CONCLUSIONS}

In this paper, we have presented a good B band photometric data of HD 40372.
We have done period analysis using a number of periodogram analysis
techniques. Final best estimate of the period is obtained by the use of
minimization of entropy method. We have determined the $Q$ value of the star
using the best determined period along with the other physical parameters
determined from the $uvby\beta$ photometry using a code developed by
Moon \& Dworetsky (1985) and later modified by Napiwotzki et al. (1993).
The $Q$ value indicates that the star is pulsating in the non-radial
p$_{3}$ mode with $l = 2$. 
 
\section{ACKNOWLEDGEMENT}
The authors would like to thank ARIES for making the telescope time available on the $104$-cm Sampurnanand telescope. SD thanks CSIR, Govt. of India for a
Senior Research Fellowship (SRF). The authors thank Dr. Mike Dworetsky for
providing the $uvby\beta$ and $tefflogg$ code. The extensive use of the SIMBAD
and ADS databases operated by the CDS center, Strasbourg, France is gratefully
acknowledged. The authors thank the anonymous referee for useful comments and suggestions.

\label{lastpage}

\end{document}